\documentclass[aps,superscriptaddress,nofootinbib,showpacs,eqsecnum,preprint,tightenlines]{revtex4}
\usepackage{hyperref}
\usepackage{epsfig,rotating}
\usepackage{amsmath,amssymb,graphicx}
\usepackage{dsfont}
\usepackage{bbm}
\numberwithin{equation}{section}

\newcommand{\be}{\begin{equation}}
\newcommand{\ee}{\end{equation}}
\newcommand{\bea}{\begin{eqnarray}}
\newcommand{\eea}{\end{eqnarray}}

\newcommand{\vx}{\vec{x}}

\newcommand{\vp}{\vec{p}}

\newcommand{\vq}{\vec{q}}

\newcommand{\vk}{\vec{k}}

\begin{document}

\title{Superhorizon entanglement  entropy\\ from particle decay in inflation.}
\author{L. Lello}
\email{lal81@pitt.edu}
\author{D. Boyanovsky}
\email{boyan@pitt.edu}
\affiliation{Department of Physics and Astronomy\\
University of Pittsburgh,
Pittsburgh PA 15260}
\author{R.~Holman}
\email{rh4a@andrew.cmu.edu}
\affiliation{Department of Physics\\
Carnegie Mellon University,
Pittsburgh PA 15213}

\date{\today}

\begin{abstract}
 In inflationary cosmology all particle states decay as a consequence of the lack of kinematic thresholds. The decay of an initial single particle state  yields an \emph{entangled quantum state of the product particles}. We generalize and extend a manifestly unitary field theoretical method   to obtain the time evolution of the quantum state.   We consider the decay of a light scalar field with mass $M\ll H$ with a cubic coupling in de Sitter space-time. Radiative corrections feature an infrared enhancement manifest as poles in $\Delta=M^2/3H^2$ and we obtain the quantum state in  an   expansion in $\Delta$. To leading order in   $\Delta$   the pure state density matrix describing the   decay of a particle with  sub-horizon wavevector is dominated by the emission of superhorizon quanta, describing \emph{entanglement between superhorizon and subhorizon fluctuations and correlations across the horizon}. Tracing over the superhorizon degrees of freedom yields a mixed state density matrix from which we obtain the entanglement entropy. Asymptotically this entropy grows with the \emph{physical} volume as a consequence of more modes of the decay products crossing the Hubble radius. A generalization to  localized wave packets is provided. The cascade decay of single particle states into many particle states is discussed.   We conjecture on \emph{possible} impact of these results on non-gaussianity and on the ``low multipole anomalies'' of the CMB.

\end{abstract}


\maketitle

\section{Introduction}
Quantum fluctuations during inflation seed the inhomogeneities which  are manifest as   anisotropies in the cosmic microwave background and primordial gravitational waves.   In its simplest inception the inflationary stage can be effectively described as a quasi-deSitter space time.  Early studies\cite{polyakov1,IR1,IR2,allen,folaci,dolgov} revealed that de Sitter space time  features infrared instabilities and profuse particle production in interacting field theories.  During inflation the rapid cosmological expansion modifies the energy-uncertainty relation allowing ``virtual'' excitations to persist longer, leading to remarkable phenomena, which is stronger in de Sitter space  time as clarified in  ref.\cite{woodard}.
 Particle production in a de Sitter background has been argued to provide a  dynamical``screening'' mechanism that leads to relaxation of the cosmological constant\cite{emil,IR3,branmore} through back reaction,  much like the production of particle-antiparticle pairs in a constant electric field. More recently this mechanism of profuse particle production   has been argued to lead to the instability of de Sitter space time\cite{polya,emil2}.

 A  particular aspect of the rapid cosmological expansion is the lack of a global time-like killing vector which leads to remarkable physical effects in de Sitter space time, as it implies the lack of kinematic thresholds (a direct consequence of energy-momentum conservation) and the decay of  fields even in their own quanta\cite{boyan,boyprem,boyquasi} with the concomitant particle production. This result that was  confirmed in ref.\cite{moschella,akhmedov,akhmerev} and more recently in ref.\cite{marolf} by a thorough analysis of the S-matrix in global de Sitter space.

 The decay of an initial single particle state into many particle states results in a quantum state that is kinematically entangled in momentum space: consider the example of a scalar field theory with cubic self-interaction and an initial single particle state with spatial physical momentum $\vec{k}$, namely $|1_{\vec{k}}\rangle$, this state decays into a two-particle states of the form $\sum_{\vec{p}} ~ C_{\vec{p}}(t)\, |1_{\vec{p}}\rangle  |1_{\vec{k}-\vec{p}}\rangle $ where $ C_{\vec{p}}(t) $ is the \emph{time dependent} amplitude of the two particle state  with  momenta $\vec{p}$ and $\vec{k}-\vec{p}$ respectively. This is an entangled state that features non-trivial correlations between the product particles. In ref.\cite{boyan,boyquasi} it is argued that in de Sitter space time with Hubble constant $H$, the largest decay amplitude corresponds to the case when   one of the product particles features physical momenta $p\ll H$, therefore, if the initial particle has physical momenta $k \gg H$ and one of the product particles features a momentum $p \ll H$ (the other with $|\vec{k}-\vec{p}|\gg H$) the quantum entangled state features correlations between the sub and superHubble daughter particles.

 We refer to these correlated pairs produced from the decay of a parent particle as   entangled across the Hubble radius, namely ``superhorizon'' entanglement, referring to the Hubble radius in de Sitter space time as the horizon as is customary in inflationary cosmology.

  Correlations of quantum fluctuations during a de Sitter inflationary stage have been recently argued\cite{masimo} to lead to remarkable Hanbury-Brown-Twiss interference phenomena with potential observational consequences.

Unitary time evolution of an initial single particle state is a pure quantum state in which the product particles are kinematically entangled.

If a pure quantum state describes an entangled state of several subsystems and if the degrees of freedom of one of the subsystems are not observed, tracing the pure state density matrix over these unobserved degrees of freedom   leads to a \emph{mixed state reduced density matrix}. The entanglement entropy is the Von Neumann entropy associated with this reduced density matrix; it reflects the loss of information that was originally present in the quantum correlations of the entangled state.

The main purpose of this article is to study the entanglement entropy in the  case of   an initial quantum state describing a single particle state with physical momentum $k \gg H$  decaying into a pair of particles one with $p\ll H$ (superhorizon), and the other with $|\vec{k}-\vec{p}|\gg H$ (subhorizon) by   tracing over the super-Hubble (``superhorizon'') degrees of freedom. This entanglement entropy is a measure of the loss of information contained in the pair correlations of the daughter particles.

 The entanglement entropy has been the focus of several studies in condensed matter systems\cite{horodecki,esentan,vedral,plenio}, statistical physics and quantum field theory\cite{cardy,srednicki,wilczek,hertzberg,huerta}, black hole physics\cite{solo,ted,enenbh} and in particle production in time dependent backgrounds\cite{hu}.  Most of  these studies focus on entanglement between spatially correlated regions across boundaries. The entanglement entropy in de Sitter space-time for a free, minimally coupled massive scalar field has been studied in ref.\cite{maldacena} with the goal of understanding superhorizon correlations,  and ref.\cite{bala} studied the entropy from momentum space entanglement and renormalization in an interacting quantum field theory in Minkowski space-time.

 Our study differs from these studies in many ways:  we are not considering spatially correlated regions, and momentum space entanglement resulting from the kinematics of particle decay in states of the same quanta is different from the cases studied in ref.\cite{bala} which considered momentum space entanglement in the interacting \emph{ground} state of two coupled theories or a finite density case, both in a stationary, equilibrium situation, whereas we are interested in the time evolution of the reduced density matrix and the concomitant increase of the entanglement entropy in an interacting theory in de Sitter space time.

 More recently the entanglement entropy in the ubiquitous case of particle decay in Minkowski space-time from tracing over the degrees of freedom of an unobserved daughter particle   has been studied in ref.\cite{minko} as a characterization of an ``invisible'' decay complementary to missing energy.

We focus on light scalar fields with mass $M^2\ll H^2$, for which  radiative corrections feature infrared divergences that are manifested as poles in $\Delta = M^2/3H^2 \ll 1$\cite{boyan,boyquasi} in the self-energy leading to a consistent expansion in $\Delta$. A similar expansion was recognized in refs.\cite{burg,rigo,leblond,smit}.

The field theoretic method introduced in ref.\cite{desiter,boyquasi,minko} that describes the \emph{non-perturbative} time evolution of quantum states is extended here and then generalized to inflationary cosmology (for other applications of this field theoretical method see refs.\cite{nuestro}) to obtain the entangled quantum state from single particle decay to leading order in a $\Delta$ expansion. We show explicitly that unitarity is manifest in the time evolution of the quantum state. From this state we construct the (pure) density matrix and trace over the contribution from superhorizon modes and obtain the entanglement entropy to leading order in a $\Delta$ expansion. Whereas in ref.\cite{2modeDS} the entanglement between \emph{only two modes} was studied in de Sitter space time, ours is a    full quantum field theoretical treatment that includes coupling between all modes as befits a local quantum field theory and consistently trace over all the superhorizon degrees of freedom.

 We find that the entanglement entropy asymptotically grows with the physical volume as more wavevectors
 cross the Hubble radius. The method is generalized to a wave packet description of single particle states and we study in detail the case of wave packets sharply localized in momentum around a wavevector $k_0 \gg H$   and localized in space on scales much smaller than the Hubble radius all throughout the near de Sitter inflationary state. We find that under these conditions, the entanglement entropy for wavepackets is approximately the  same as that for plane waves and assess the corrections.

 As mentioned above, the lack of kinematic thresholds implies that quanta can decay on many quanta of the same field, in particular for cubic interactions a single particle state can decay into two particles of the same field, however the decay process does not stop at the two particle level, but instead is a
 \emph{cascade} decay $1\rightarrow 2\rightarrow 3 \rightarrow \cdots$. We provide a non-perturbative framework    to study this cascade decay process and argue that for weak (cubic) coupling $\lambda$  there is a hierarchy of time scales and the cascade is controlled by this weak coupling. The probability of multiparticle   states is suppressed by $\lambda^2$ for each extra particle in the final state, the time scales of production and decay of multiparticle states are also separated by $1/\lambda^2$.

 We comment on possible relationship with non-gaussianity, in particular pointing out the relationship between the quantum correlations between subhorizon and superhorizon quanta from particle decay and the bispectrum of scalar perturbations in the squeezed (local) limit. Furthermore, we speculate as to whether the information ``lost'' as modes cross the horizon is ``recovered'' when the modes re-enter the horizon during the matter dominated era. This study then bridges the main concepts of entanglement between spatial regions explored in ref.\cite{maldacena}, with momentum space entanglement and coarse graining\cite{bala} and quantum entanglement via particle decay\cite{minko}  in inflationary cosmology.

\section{Quantum Field Theoretical Wigner-Weisskopf treatment of the decay width}
The method developed in refs.\cite{desiter,boyquasi,minko,nuestro} is a quantum field theoretical generalization of the Wigner-Weisskopf method used in quantum optics\cite{ww,books}.

We consider a scalar field minimally coupled to gravity in a spatially flat  de Sitter spacetime with scale factor $a(t)=e^{Ht}$ . In comoving
coordinates, the action is given by
\begin{equation}
S= \int d^3x \; dt \;  a^3(t) \Bigg\{
\frac{1}{2}{\dot{\phi}^2}-\frac{(\nabla
\phi)^2}{2a^2}-\frac{M^2}{2} \phi^2
- \lambda \;  \phi^{\,3}  \Bigg\},,\label{lagrads}
\end{equation}

It is convenient to pass to conformal time $\eta = -e^{-Ht}/H$ with $d\eta =
dt/a(t)$ and introduce a conformal rescaling of the fields
\begin{equation}
a(t)\phi(\vx,t) = \chi(\vx,\eta).\label{rescale}
\end{equation}
The action becomes (after discarding surface terms that will not
change the equations of motion) \be S  =
  \int d^3x \; d\eta  \; \Bigg\{\frac12\left[
{\chi'}^2-(\nabla \chi)^2-\mathcal{M}^2 (\eta) \; \chi^2  \right] -\lambda \, C (\eta)  \;  \chi^3   \Bigg\} \; , \label{rescalagds}\ee
with primes denoting derivatives with respect to
conformal time $\eta$ and \be \mathcal{M}^2 (\eta) =
M^2C^2(\eta)-\frac{C''(\eta)}{C(\eta)}  \; , \label{massds}
\ee
where for de Sitter spacetime \be C(\eta)= a(t(\eta))= -\frac{1}{H\eta}. \label{scalefactords}\ee  In this case, the effective time dependent mass is given by
\be
\mathcal{M}^2 (\eta)  = \Big[\frac{M^2 }{H^2}-2\Big]\frac{1}{\eta^2}   \; . \label{massds2}
\ee
The free field Heisenberg equations of motion for the spatial Fourier modes of the field with wavevector $k$  are given by
\be  \chi''_{\vk}(\eta)+
\Big[k^2-\frac{1}{\eta^2}\Big(\nu^2 -\frac{1}{4} \Big)
\Big]\chi_{\vk}(\eta)  =   0   \label{dsmodes}  \; ,
\ee
\noindent where
\be
\nu^2   =  \frac{9}{4}- \frac{M^2}{H^2}\,.
   \label{nu}\ee
This can be solved to find the two linearly independent solutions of (\ref{dsmodes}):
\bea
g_{\nu}(k;\eta) & = & \frac{1}{2}\; i^{\nu+\frac{1}{2}}
\sqrt{-\pi \eta}\,H^{(1)}_\nu(-k\eta)\label{gnu}\\
f_{\nu}(k;\eta) & = & \frac{1}{2}\; i^{-\nu-\frac{1}{2}}
\sqrt{-\pi \eta}\,H^{(2)}_\nu(-k\eta)= g^*_{\nu}(k;\eta) \label{fnu}  \; ,
\eea
 \noindent where $H^{(1,2)}_\nu(z)$ are Hankel functions. Expanding the field operator in this basis yields
\be \chi(\vec{x},\eta) = \frac{1}{\sqrt{V}}\sum_{\vec{k}} \Big[a_{\vec{k}}\,g_\nu(k;\eta)\,e^{i\vec{k}\cdot\vec{x}}+ a^\dagger_{\vec{k}}\,\,g^*_\nu(k;\eta)\,e^{-i\vec{k}\cdot\vec{x}}\Big]\,. \label{quantumfieldds1}\ee
The Bunch-Davies vacuum is defined such that \be a_{\vec{k}}|0\rangle = 0 \,,\label{dsvac}\ee and the Fock space states are obtained in the usual manner, i.e. by applying creation operators $a_{\vec{k}}^{\dagger}$ to the vacuum.

In what follows we consider a light scalar field with $M\ll H$ and write
\be \nu = \frac{3}{2}-\Delta ,\ \Delta = \frac{M^2}{3H^2} +\cdots \ll 1\,.  \label{nudefi}\ee For light scalar fields with $\Delta \ll 1$ quantum loop corrections feature an infrared enhancement from the emission and absorption of superhorizon quanta that is manifest as poles in $\Delta$\cite{boyan,boyquasi}. Below we exploit the expansion in $\Delta$ implemented in ref.\cite{boyan,boyquasi} to leading order, isolating the most infrared sensitive contributions to the entanglement entropy from these processes.

 In the Schr\"oedinger picture the quantum states $|\Psi(\eta)\rangle$ obey
 \be i\frac{d}{d\eta}|\Psi(\eta)\rangle = H(\eta) \, |\Psi(\eta)\rangle \label{Spic}\ee where in an expanding cosmology the Hamiltonian $H(\eta)$ is generally a function of $\eta$ in marked contrast to the situation in Minkowski space-time, where it is constant. Introducing the time evolution operator $U(\eta,\eta_0)$ obeying
 \be i\frac{d}{d\eta} U(\eta,\eta_0) = H(\eta)\,U(\eta,\eta_0) , \quad U(\eta_0,\eta_0) = 1, \label{Uds}\ee the solution of the Schr\"oedinger equation is $|\Psi(\eta)\rangle = U(\eta,\eta_0)\,|\Psi(\eta_0)\rangle $. Now separate out the interaction Hamiltonian by writing $H(\eta) = H_0(\eta) + H_{i}(\eta)$ with $H_0(\eta)$ the non-interacting Hamiltonian, and introduce the time evolution operator of the free theory $U_0(\eta,\eta_0)$ satisfying
 \be i\frac{d}{d\eta} U_0(\eta,\eta_0) = H_0(\eta)\, U_0(\eta,\eta_0), \quad i\frac{d}{d\eta} U^{-1}_0(\eta,\eta_0) = - U^{-1}_0(\eta,\eta_0)\, H_0(\eta) , \quad U_0(\eta_0,\eta_0) =1, \label{U0ds}\ee the interaction picture states are defined as
 \be |\Psi(\eta)\rangle_I = U_I(\eta,\eta_0)|\Psi(\eta_0)\rangle_I =  U^{-1}_0(\eta,\eta_0) |\Psi(\eta)\rangle. \label{ipds}\ee Here $U_I(\eta,\eta_0)$ is the time evolution operator in the interaction picture and obeys
 \be  \frac{d}{d\eta}U_I(\eta,\eta_0) = -i H_I(\eta) U_I(\eta,\eta_0), \quad U_I(\eta_0,\eta_0)=1 \label{UI}\ee  and \be H_I(\eta) = U^{-1}_0(\eta,\eta_0) H_{i}(\eta) U_0(\eta,\eta_0), \label{HIdsdef}\ee  where $\chi$ is the free field Heisenberg field operator in eq.(\ref{quantumfieldds1}).

\subsection{Transition amplitudes and probability}
Now consider a cubic interaction Hamiltonian for a scalar field which we label as $\chi(\vec{x},\eta)$ after the conformal rescaling described above:
\be H_I(\eta) = -\frac{\lambda}{H\,\eta} \int d^3 x ~ \chi^3(\vec{x},\eta) \,. \label{gencub}\ee  We can then use the expansion of the scalar field $\chi$ given by (\ref{quantumfieldds1}) to compute the transition amplitude for a one particle state to decay into two particles $\chi_{\vk}  \rightarrow  \chi_{\vp}+\chi_{\vk-\vp}$  as depicted in fig. (\ref{fig:decay}):
\begin{figure}[h!]
\includegraphics[keepaspectratio=true,width=2in,height=2in]{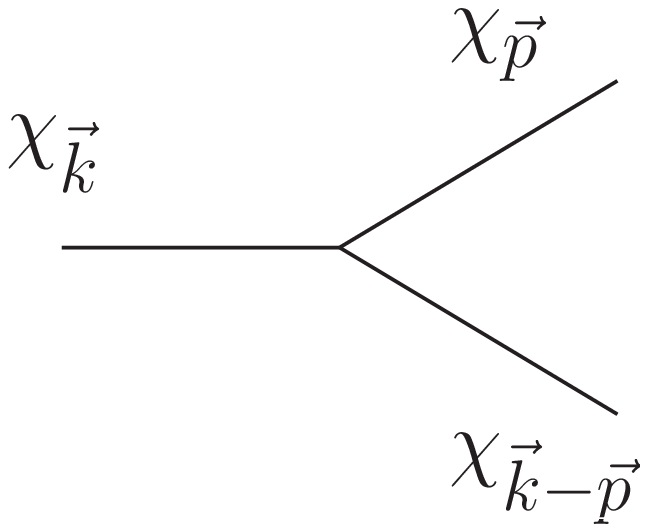}
\caption{The decay  $\chi_{\vk}  \rightarrow  \chi_{\vp}+\chi_{\vk-\vp}$.}
\label{fig:decay}
\end{figure}

\be \mathcal{A}_{\chi \rightarrow \chi\chi}(\vec{k},\vec{p};\eta) = \frac{6 \, i \,\lambda}{H \,\sqrt{V}} \int^\eta_{\eta_0} \frac{d\eta_1}{\eta_1}\,g_\nu(k;\eta_1)\,g^*_\nu(p;\eta_1)\,g^*_\nu(|\vec{k}-\vec{p}|;\eta_1).  \label{amp}\ee The total transition probability is
\be \mathcal{P}_{\chi \rightarrow \chi\chi}(k;\eta) =  V \int \frac{d^3p}{(2\pi)^3} ~\big| \mathcal{A}_{\chi \rightarrow \chi\chi}(\vec{k},\vec{p};\eta)\big|^2=  \int^\eta_{\eta_0}  {d\eta_2}  \int^\eta_{\eta_0} {d\eta_1} ~\Sigma(k\,;\eta_1,\eta_2) \label{proba}\ee where
\be \Sigma(k\,;\eta_1,\eta_2) = \frac{36 \,\lambda^2~g^*_\nu(k,\eta_2)\,g_\nu(k,\eta_1)}{H^2 \,\eta_1\,\eta_2} \,\int \frac{d^3p}{(2\pi)^3} ~ g^*_{\nu}(p,\eta_1)\,g^*_{\nu}(q,\eta_1)\,g_{\nu}(p,\eta_2)\,g_{\nu}(q,\eta_2),  \label{selfe}\ee  where $ q=|\vec{k}-\vec{p}|$. Note that this kernel has the property that
\be \Sigma(k\,;\eta_2,\eta_1) = \Sigma^*(k\,;\eta_1,\eta_2)\,. \label{proper}\ee
Introducing the identity $1 = \Theta(\eta_2-\eta_1)+\Theta(\eta_1-\eta_2)$ in the (conformal) time integrals and using (\ref{proper}) we find
\be \mathcal{P}_{\chi \rightarrow \chi\chi}(k;\eta) = 2   \int^\eta_{\eta_0}  {d\eta_2}  \int^{\eta_2}_{\eta_0}  {d\eta_1} ~\mathrm{Re}\Big[\Sigma(k\,;\eta_1,\eta_2) \Big] \label{proba2}\ee from which we obtain the \emph{transition rate} as
\be \Gamma(\eta) \equiv  \frac{d}{d\eta} \mathcal{P}_{\chi \rightarrow \chi\chi}(k;\eta) =  2   \int^\eta_{\eta_0}  {d\eta'}   ~\mathrm{Re}\big[\Sigma(k\,;\eta ,\eta') \big] \label{gamma}\ee

In Minkowski space-time ($\eta \rightarrow t$),  if the kinematics of the transition is allowed, i.e. if energy-momentum conservation obtains, the transition is to on-shell states and the transition probability grows linearly in time, exhibiting  secular growth. In the long time limit the transition rate becomes a constant. This is basically how the result from Fermi's Golden rule comes about. If, on the other hand energy-momentum conservation is not fulfilled, the probability becomes constant at asymptotically long times, with a vanishing transition rate, describing virtual processes that contribute to wave function renormalization. A true decay of the quantum state is therefore reflected in secular \emph{growth} of the transition probability and a transition rate that either remains constant or grows at asymptotically long time. In de Sitter space time the lack of a global time-like Killing vector implies the lack of kinematic thresholds. As discussed earlier in ref.\cite{boyquasi,boyan,desiter} and confirmed in ref.\cite{marolf}, quanta of a single field can decay into other quanta of the same field regardless of the mass of the field.

\subsection{Wigner-Weisskopf theory in de Sitter space time:}\label{subsec:ww}

In this subsection, we review the work in refs.\cite{desiter,boyquasi,minko,nuestro}, as the implementation of the quantum field theoretical Wigner-Weisskopf formulation is crucial in constructing states whose time evolution is manifestly unitary.

 Expanding the interaction picture state $|\Psi(\eta)\rangle_I$ in   Fock states $|n\rangle$ obtained as usual by applying the creation operators on to the (bare) vacuum state as
  \be |\Psi(\eta)\rangle   = \sum_n C_n(\eta) |n\rangle \label{expastate}\ee the evolution of the state in the interaction picture given by eqn. (\ref{ipds}) yields
  \be i \frac{d}{d\eta}|\Psi(\eta)\rangle  = H_I(\eta)|\Psi(\eta)\rangle  \label{eomip}\ee which in terms of the coefficients $C_n(\eta)$ become
   \be \frac{d\,C_n(\eta)}{d\eta}  = -i \sum_m C_m(\eta) \langle n|H_I(\eta)|m\rangle \,, \label{ecns}\ee it is convenient to separate the diagonal matrix elements from those that represent transitions, writing
    \be \frac{d\,C_n(\eta)}{d\eta}  = -i C_n(\eta)\langle n|H_I(\eta)|n\rangle -i \sum_{m\neq n} C_m(\eta) \langle n|H_I(\eta)|m\rangle \,. \label{ecnsoff}\ee
    Although this equation is exact, it provides an infinite hierarchy of simultaneous equations when the Hilbert space of states $|n\rangle$ is infinite dimensional.
    %%%n new addition #1
    The Wigner-Weisskopf method consists of two main ingredients: i) truncation of the hierarchy at a given order in the perturbative expansion, ii) a Markovian approximation that yields the long time asymptotics of the coefficients.

    In ref.\cite{desiter} the equivalence between the Wigner-Weisskopf method, the time evolution obtained from the Dyson resummation of propagators in terms of the self-energy and the dynamical renormalization group was shown in Minkowski space time. Hence this method provides a non-perturbative resummation  to obtain the real time dynamics  of quantum states.
    %% end of new addition #1.

   %% new addition #2

   We begin by implementing this program to lowest order, and provide a roadmap for implementation at arbitrary higher order in section (\ref{sec:cascade}) where we also study ``cascade processes'' that are available in de Sitter space time.

   %% end of new addition #2

    Thus,
consider the case when a state $|A\rangle$, say, couples to a set of states $|\kappa\rangle$, which in turn couple back to $|A\rangle$ via $H_I$. Then to lowest order in the interaction, the system of equation closes in the form
 \bea \frac{d\,C_A(\eta)}{d\eta} & = & -i   \langle A|H_I(\eta)|A\rangle \, C_A(\eta)-i \sum_{\kappa \neq A} \langle A|H_I(\eta)|\kappa\rangle \,C_\kappa(\eta)\label{CA}\\
\frac{d\,C_\kappa(\eta)}{d\eta}& = & -i   \langle \kappa|H_I(\eta)|\kappa\rangle \, C_\kappa(\eta)-i \,   \langle \kappa|H_I(\eta) |A\rangle \, C_A(\eta)\label{Ckapas}\eea where the $\sum_{\kappa \neq A}$ is over all the intermediate states coupled to $|A\rangle$ via $H_I$ representing transitions. By including the diagonal terms $\langle n|H_I(\eta)|n\rangle\,C_n$ specifically, we can also consider mass counterterms\cite{boyquasi}, however, we will neglect these terms in the sequel since we are not concerned with either mass generation or renormalization in this article.

Consider the initial value problem in which at time $\eta=\eta_0$ the state of the system is given by $|\Psi(\eta=\eta_0)\rangle = |A\rangle$ so that \be C_A(\eta_0)= 1\ , C_{\kappa \neq A}(\eta=\eta_0) =0.\label{initial}\ee  We can then solve (\ref{Ckapas}) and substitute the solution back into (\ref{CA}) to find \bea  C_{\kappa}(\eta) & = &  -i \,\int_{\eta_0}^{\eta} \langle \kappa |H_I(\eta')|A\rangle \,C_A(\eta')\,d\eta' \label{Ckapasol}\\ \frac{d\,C_A(\eta)}{d\eta} & = &   - \int^{\eta}_{\eta_0} \Sigma(\eta,\eta') \, C_A(\eta')\,d\eta' \label{intdiff} \eea where \be \Sigma(\eta,\eta') = \sum_{\kappa } \langle A|H_I(\eta)|\kappa\rangle \langle \kappa|H_I(\eta')|A\rangle. \label{sigma} \ee

This integro-differential equation with \emph{memory} yields a non-perturbative solution for the time evolution of the amplitudes and probabilities. We can construct an approximation scheme to solve this equation as follows. First note that the time evolution of $C_A(\eta)$ as determined by eqn. (\ref{intdiff}) is \emph{slow} in the sense that
the relevant time scale is determined by a weak coupling kernel $\Sigma$. This allows us to introduce a Markovian approximation in terms of an expansion in derivatives of $C_A$ as follows: define
\be W_0(\eta,\eta') = \int^{\eta'}_{\eta_0} \Sigma(\eta,\eta'')d\eta'' \label{Wo}\ee so that \be \Sigma(\eta,\eta') = \frac{d}{d\eta'}\,W_0(\eta,\eta'),\quad W_0(\eta,\eta_0)=0. \label{rela} \ee Integrating by parts in eq.(\ref{intdiff}) we obtain \be \int_{\eta_0}^{\eta} \Sigma(\eta,\eta')\,C_A(\eta')\, d\eta' = W_0(\eta,\eta)\,C_A(\eta) - \int_{\eta_0}^{\eta} W_0(\eta,\eta')\, \frac{d}{d\eta'}C_A(\eta') \,d\eta'. \label{marko1}\ee The first term has ``erased'' the memory in the kernel by setting both time arguments to be the time of interest, while the second term on the right hand side is formally of \emph{fourth order} in $H_I$.  Integrating by parts successively as discussed in ref.\cite{desiter} a systematic approximation scheme can be developed. To leading order in the coupling (second order in $H_I$), we will neglect the second term on the right hand side of (\ref{marko1}), in which case eqn. (\ref{intdiff}) becomes
\be \frac{d\,C_A(\eta)}{d\eta} + W_0(\eta,\eta)\,C_A(\eta) =0 \label{markoeA}\ee with solution
 \be C_A(\eta) = e^{-\int^{\eta}_{\eta_0} {W}_0(\eta',\eta')\, d\eta'}  , \quad  {W}_0(\eta',\eta')=   \int_{\eta_0}^{\eta'} \Sigma(\eta',\eta^{''}) d\eta^{''}\,. \label{dssolu} \ee Introducing the \emph{real quantities} $\mathcal{E}_A(\eta), \ \Gamma_A(\eta)$ as
 \be   \int^{\eta'}_{\eta_0} \Sigma(\eta',\eta'')d\eta'' = i\,\mathcal{E}_A(\eta')+ \frac{1}{2}~\Gamma_A(\eta') \label{realima}\ee  where
 \be \Gamma_A(\eta') = 2 \int^{\eta'}_{\eta_0} \mathrm{Re}\Big[\Sigma(\eta',\eta'')\Big] d\eta'' \label{gama2}\ee
 in terms of which
 \be  C_A(\eta) = e^{-i\int^{\eta}_{\eta_0}\mathcal{E}_A(\eta') d\eta'}~ e^{-\frac{1}{2}\int^{\eta}_{\eta_0}\Gamma_A(\eta') d\eta'}\,. \label{caofeta}\ee When the state $A$ is a single particle state, radiative corrections to the mass are extracted from $\mathcal{E}_A$ and
 \be \Gamma_A(\eta) = -  \frac{d}{d\eta}\ln\Big[\big|C_A(\eta)\big|^2\Big]  \label{decarate} \ee is identified as a (conformal) time dependent decay rate. Comparing these expressions with the transition probability (\ref{proba2}) we see from (\ref{decarate}) that
 \be \big|C_A(\eta)\big|^2 = e^{-\mathcal{P}_{\chi \rightarrow \chi\chi}(k;\eta)} \,,\label{equiva} \ee and that $\Gamma(\eta)$ is exactly the same as expression (\ref{gamma}).
\subsection{Unitarity}
One of our main goals is to study the entanglement entropy from tracing over superhorizon degrees of freedom. Thus
it is important to make sure that the loss of information encoded in the entanglement entropy is a genuine effect of the tracing procedure and not a consequence of approximations in the evolution of the quantum state.  Unitarity follows from the set of equations (\ref{eomip}), combining these with their complex conjugates it is straightforward to confirm that
\be \frac{d}{d\eta} \sum_{n} |C_n(\eta)|^2 = 0\,.\label{constsum}\ee therefore with the initial conditions (\ref{initial}) it follows that
\be \sum_{n} |C_n(\eta)|^2 =1\,.  \label{unione}\ee

Although this is an exact statement,  we now show that the Wigner-Weisskopf approximation and its Markovian implementation maintain unitary time evolution.

Using (\ref{Ckapasol}) consider
\be \sum_{\kappa}|C_\kappa(\eta)|^2 = \int_{\eta_0}^{\eta}d\eta_1 C^*_A(\eta_1)\int_{\eta_0}^{\eta}d\eta_2\Sigma(\eta_1,\eta_2) C_A(\eta_2). \label{sumkapa}\ee Inserting $1=\Theta(\eta_1-\eta_2)+\Theta(\eta_2-\eta_1)$ as we did earlier,  it follows that
\bea \sum_{\kappa}|C_\kappa(\eta)|^2 & = &  \int_{\eta_0}^{\eta}d\eta_1 C^*_A(\eta_1)\int_{\eta_0}^{\eta_1}d\eta_2\Sigma(\eta_1,\eta_2) C_A(\eta_2)  \nonumber \\ & + & \int_{\eta_0}^{\eta}d\eta_2 C_A(\eta_2)\int_{\eta_0}^{\eta_2}d\eta_1\Sigma(\eta_1,\eta_2) C^*_A(\eta_1).\label{shuffle} \eea Using $\Sigma (\eta_1,\eta_2) = \Sigma^*(\eta_2,\eta_1)$, relabelling $\eta_1 \leftrightarrow \eta_2$ in the second line of (\ref{shuffle}) and using (\ref{intdiff}), we find
\bea  \sum_{\kappa}|C_\kappa(\eta)|^2  &=& - \int_{\eta_0}^{\eta}d\eta_1 \Big[ C^*_A(\eta_1)\frac{d}{d \eta_1}C_A(\eta_1) + C_A(\eta_1) \frac{d}{d \eta_1}C^*_A(\eta_1)\Big]\nonumber\\
&=& - \int_{\eta_0}^{\eta}d\eta_1 \frac{d}{d\eta_1} |C_A(\eta_1)|^2 = 1-|C_A(\eta)|^2 \label{unita}\eea where we have used the initial condition $C_A(\eta_0)=1$. This is the statement of unitary time evolution, namely
\be  |C_A(\eta)|^2 + \sum_{\kappa}|C_\kappa(\eta)|^2 = |C_A(\eta_0)|^2 \label{unitime}\ee
To leading order in the Markovian approximation, the unitarity relation becomes
\be  \sum_{\kappa}|C_\kappa(\eta)|^2   =   -2 \int_{\eta_0}^{\eta}  \Big|C_A(\eta_1)\Big|^2 \,\mathrm{Re}\Big[W_0(\eta_1,\eta_1)\Big] \,d\eta_1  =  1-|C_A(\eta)|^2 \label{unimark}\ee where $C_A(\eta_0)=1$.

\section{Particle decay: entanglement across the horizon:}
In the scalar theory described by eq.(\ref{gencub}) the cubic interaction allows a single particle state $|1_{\vk}\rangle$ to  \emph{decay} into two particle states $|1_{\vk-\vp};1_{\vp}\rangle$\cite{boyan,boyquasi}. To lowest order in the coupling  the matrix element for this process is given up to an overall phase by
\be \mathcal{M}(p;k;\eta)= \langle 1_{\vk-\vp};1_{\vp}|H_I(\eta)|1_{\vk}\rangle = -\frac{6\lambda}{H\eta\sqrt{V}} \, g_\nu(k;\eta)\,g^*_{\nu }(p;\eta)\,g^*_{\nu }(|\vec{k}-\vec{p}|;\eta)\,.\label{mtxele1}\ee Consider an initial single particle state $|1_{\vk}\rangle$ at time $\eta_0$. Upon time evolution in the interaction picture this state evolves into
\be |\Psi(\eta)\rangle_I  = C_{k}(\eta)|1_{\vk}\rangle + \sum_{\vp} C_{p}(k;\eta) |1_{\vk-\vp};1_{\vp}\rangle~~;~~ C_k(\eta_0)=1~;~C_{p }(k,\eta_0)=0\,. \label{psi}\ee This is an \emph{entangled} state in which pairs of particles with momenta $\vk-\vp,\ \vp$ \emph{are correlated}. In particular if $\vk$ is \emph{subhorizon} and $\vp$ is \emph{superhorizon}, the quantum state (\ref{psi}) describes entanglement and correlation of particles across the horizon.

The coefficients in the state (\ref{psi}) are the solutions of the (WW) equations, namely
\be \frac{d}{d\eta}\,C_k(\eta)   =   -\int^{\eta}_{\eta_0} \Sigma(k,\eta,\eta')\,C_k(\eta')\,d\eta' \,, \label{Ckww}\ee
 \be C_{p}(k;\eta) = -i \int^{\eta}_{\eta_0} \mathcal{M}(p;k;\eta')\, C_k(\eta')\, d\eta' \,,\label{coefpk}\ee where the matrix element is given by eq.(\ref{mtxele1}). We will focus on the asymptotic limit where $\eta\rightarrow 0^-;\eta_0 \rightarrow -\infty$.

 The self-energy     (\ref{sigma})  is given by\footnote{This expression corrects a prefactor in ref.\cite{boyquasi}.}
\be    \Sigma(k,\eta,\eta')   =  \sum_{\vp} \mathcal{M}^*(p;k;\eta)\,\mathcal{M}(p;k;\eta') \,,
\label{sefull}\ee  were the matrix elements are given by eq.(\ref{mtxele1}) leading to the result given by (\ref{selfe}).

As discussed in detail in ref.\cite{boyquasi}, as $\Delta \rightarrow 0$  the integral features infrared divergences from regions in which the momenta are superhorizon, namely $p\eta,p\eta' \ll 1$ and  $|\vk-\vp|\eta, |\vk-\vp|\eta' \ll 1$. Both of these momentum regions yield the same infrared contribution as a single pole in $\Delta$\cite{boyquasi}, as can be seen as follows. For superhorizon modes ($-p\eta \ll1$) the mode functions (\ref{gnu}) behave (up to an overall phase) as
\be g_\nu(p;\eta) \simeq \frac{1}{\sqrt{2}}\frac{1}{p^{\frac{3}{2}-\Delta}\,(-\eta)^{1-\Delta}} \label{superho}\ee and for subhorizon modes $-k\eta \gg 1$
\be g_\nu(k;\eta) = \frac{1}{\sqrt{2k}}~e^{-ik\eta}.\   \label{subho}\ee Therefore for $p\ll (-1/\eta)$ and $k \gg (-1/\eta)$ the matrix element (\ref{mtxele1}) becomes (up to an overall phase)
\be \mathcal{M}(p,k;\eta) \simeq  \frac{6\,\lambda}{2\sqrt{2} Hk \sqrt{V}\,(-\eta)^{2-\Delta} } \,\frac{1}{p^{\frac{3}{2}-\Delta}}.\label{matsub}\ee The contribution to the self-energy from superhorizon modes with $p \leq \mu  \lesssim (-1/\eta)$ (with $\mu$ an infrared cutoff) yields
\be  \frac{V}{2\pi^2}\int^\mu_0 p^2\,\mathcal{M}^*(p;k;\eta)\,\mathcal{M}(p;k;\eta') \,dp   =   \frac{9\lambda^2}{8\pi^2 H^2 k^2 \eta^2 {\eta'}^2 \Delta}\,\big[1+\Delta \ln[\mu^2 \eta \eta']+\cdots \big] \,. \label{integ1}\ee

\begin{figure}[ht!]
\begin{center}
\includegraphics[height=4in,width=4in,keepaspectratio=true]{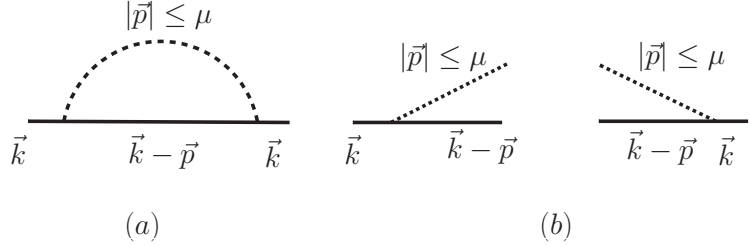}
\caption{Processes that contribute to the leading order poles in $\Delta$: (a) intermediate state of superhorizon modes, (b) emission and absorption of superhorizon quanta, with $\mu \lesssim (-1/\eta)$.}
\label{figsoftse}
\end{center}
\end{figure}

The processes that contributes to leading order in $\Delta$ is the emission of \emph{superhorizon quanta}, depicted in fig. (\ref{figsoftse})

 The simple rules to extract the leading order contribution in $\Delta$ are given in ref.\cite{boyquasi}, where the cancellation of the infrared regulator $\mu$  from the contributions of the subhorizon modes, for which one can safely set $\Delta =0$,  is also shown in detail. In particular, the appendix of ref.\cite{boyquasi} shows how the contribution of the subhorizon modes replaces the term $\ln[\mu^2 \eta\eta'] \rightarrow \ln[k^2\eta\eta']$ which to leading order in $\Delta$  can be written as  $1 + \Delta \ln[k^2\eta\eta']\simeq [k^2\eta\eta']^{\Delta}$. The contribution from the region $|\vk-\vp| \ll \mu$ yields an overall factor $2$ in the self-energy so that to leading order in $\Delta$ (as can be seen by rerouting the loop momentum) \be \Sigma(k,\eta,\eta') = \frac{\alpha}{k^{2-2\Delta}\eta^{2-\Delta}\,\eta^{'\,2-\Delta}}~~,~~  \alpha= \frac{9\,\lambda^2}{4\pi^2\, H^2\,\Delta }= \frac{27\,\lambda^2}{4\pi^2\, M^2  }\, . \label{sigfina}\ee

Using the result in eq.(\ref{dssolu}) we finally find that to leading order,
\be C_k(\eta) = \exp{\Big[-\frac{\alpha}{2\,z^{(2-2\Delta)}}\Big]},\ ~z=(-k\eta)\,, \label{Ckfinal} \ee where we have approximated $\alpha/2\,z^{2-2\Delta}_0 \rightarrow 0$ since $-k \eta_0 \gg 1$ as the physical wavevector of the initial particle is deep inside the Hubble radius at the initial time  and it is assumed to remain inside the Hubble radius during the evolution.

\section{Entanglement entropy}

The pure state density matrix corresponding to the entangled state of eq.(\ref{psi}) is
\be \rho(\eta) = |\Psi(\eta)\rangle \langle \Psi(\eta)|. \label{rho}\ee Now let us trace over the superhorizon physical wavevectors $-\vp \eta\lesssim 1$. This leads us to the \emph{mixed state} density matrix for modes whose wavelengths are \emph{inside} the horizon during the evolution
\be \rho_r(\eta) = |C_k(\eta)|^2 |1_{\vk}\rangle \langle 1_{\vk}| + 2 \sum_{-p\eta \lesssim 1} |C_{p}(k;\eta)|^2
|1_{\vk-\vp}\rangle \langle 1_{\vk-\vp}| \label{rhored}\ee where the factor $2$ accounts for the two regions of superhorizon (physical) momenta $-p\eta < 1$ and $-|\vk-\vp|\eta< 1$ which yield the same contribution, as can be easily seen after a relabelling of momenta.

The entanglement entropy is the Von-Neumann entropy for the reduced density matrix, we find
\be S(\eta) = -n_k(\eta)\ln n_k(\eta) - 2\sum_{p \lesssim (-1/\eta)} n_p(\eta) \ln n_p(\eta) \label{entropy} \ee
where the occupation numbers of the initial and \emph{produced} quanta are given by
\be n_k(\eta) = \langle \Psi(\eta)|a^\dagger_{\vk}\, a_{\vk}|\Psi(\eta)\rangle = |C_k(\eta)|^2,\ n_p(\eta) = \langle \Psi(\eta)|a^\dagger_{\vp}\, a_{\vp}|\Psi(\eta)\rangle = |C_p(k;\eta)|^2 \,.\label{ocupa}\ee Note that the unitarity relation in eq.(\ref{unimark}) implies that
\be \sum_{\vp} n_p(\eta) = 1- n_k(\eta)\,. \label{unitanumber}\ee as expected on physical grounds.

At this stage it is important to highlight how unitarity is manifest to leading order in the $\Delta$ expansion  as this  feature  simplifies the calculation of the entanglement entropy considerably. The main point is that the unitarity relation (\ref{unimark}) implies that the contribution of superhorizon modes is the dominant one. This can be seen clearly from the following arguments: to leading order in $\Delta$ we can neglect the term $2\Delta$ in the exponent of $z$ in the solution (\ref{Ckfinal}) and in the term $-\Delta$ in the exponent of $(-\eta)$ in the matrix element (\ref{matsub}) for superhorizon modes. Now the coefficient
\be C_p(k;\eta) = -i \int^{\eta}_{\eta_0} \mathcal{M}(p;k;\eta_1)\, C_k(\eta_1) \,d\eta_1 \simeq -i \frac{2\pi} {\sqrt{2V}}\,\frac{\sqrt{\Delta}}{p^{\frac{3}{2}-\Delta}}~\int^{y(\eta)}_{y(\eta_0)} e^{-y^2/2} dy \label{coef}\ee where we used the definition of $\alpha$ given by eqn. (\ref{sigfina}) and changed variables of integration to $\eta_1 = \sqrt{\alpha}/k\, y$. This expression clearly exhibits that the contribution to $|C_p(k;\eta)|^2$ from superhorizon modes,  to leading order in $\Delta$ can be written in the following factorized form: \be |C_{p}(k;\eta)|^2 =  F[k;\eta]~ \frac{\Delta}{V\,p^{3-2\Delta}}\,. \label{ansat}\ee The dependence on $\Delta$ is a manifestation of unitarity to leading order; if we compute the integral in  eq.(\ref{ansat}) over superhorizon modes
 \be \sum_{p \lesssim (-1/\eta)}|C_{p}(k;\eta)|^2 =  \frac{ F[k;\eta]\,\Delta}{2\pi^2}~ \int^{(-1/\eta)}_0 \frac{p^2 dp}{p^{3-2\Delta}} =   \frac{F[k;\eta]}{4\pi^2}\,(-1/\eta)^{2\Delta}, \label{inteuni}\ee  the $\Delta$ in the numerator in eq.(\ref{ansat}) cancels the single pole in $\Delta$ from the integral giving an $\mathcal{O}(1)$ contribution, which is what is necessary to satisfy the unitarity condition (\ref{unimark}) to leading order in $\Delta$.

 This result is similar to that found in the case of particle decay in Minkowski space time\cite{minko}: in this case the particles produced from the decay of a parent particle feature a Lorentzian distribution in energy, with width $\Gamma$ the decay width of the parent particle and amplitude $1/\Gamma$, so that the energy integral over the distribution is $\mathcal{O}(1)$. In ref.\cite{minko} it is proven to leading order in the perturbative expansion $\mathcal{O}(\Gamma)$ that this narrow distribution of large amplitude is the main reason for the fulfillment of unitarity to leading order in the Wigner-Weisskopf approximation. In the case of de Sitter space time, the distribution function of the particles produced with superhorizon wavevectors is $\propto \Delta /p^{3-2\Delta}$ whose momentum integral over the region of superhorizon momenta is also of $\mathcal{O}(1)$.

 Thus in the limit $\Delta \ll 1$ the sum $\sum_{p}\,|C_p(\eta)|^2$ is dominated by the superhorizon momenta and from the unitarity relation (\ref{unimark}) we find
 \be \mathrm{Tr} \rho_r(\eta) = |C_k(\eta)|^2 + \sum_p\,|C_p(\eta)|^2 =1  \,.\label{trace1} \ee

Although the integral in $F[k;\eta]$ can be written in terms of error functions, the unitarity relation (\ref{unimark}) and the result (\ref{trace1}) furnish  a more direct evaluation. Consider
\bea \sum_{p} |C_{p}(k;\eta)|^2 & = &  \int^\eta_{\eta_0}  \int^\eta_{\eta_0}   \sum_{ p} \mathcal{M}^*(p;k;\eta_1)\,\mathcal{M}(p;k;\eta_2)\, C^*_k(\eta_1)\,C_k(\eta_2)\,d\eta_1 \, d\eta_2 \nonumber \\ &  = &   \int^\eta_{\eta_0}  \int^\eta_{\eta_0}   \Sigma(k,\eta_1,\eta_2) \, C^*_k(\eta_1)C_k(\eta_2)\,d\eta_1 \, d\eta_2 \,.\label{sumita}\eea This is the same expression as in eqn. (\ref{sumkapa}), so that implementing the same steps as in eqns. (\ref{shuffle},\ref{unita}) leads to the unitarity relation (\ref{unimark}), namely
\be  \sum_{p} |C_{p}(k;\eta)|^2 = 1- |C_k(\eta)|^2 \,.\label{unifini}\ee To leading order in $\Delta$, the sum is dominated by the superhorizon contributions from both regions of integrations $p \lesssim (-1/\eta)~,~|\vk-\vp|\lesssim (-1/\eta)$ contributing equally, hence
 \be  \sum_{p\lesssim (-1/\eta)} |C_{p}(k;\eta)|^2 \simeq \frac{1}{2} \big[ 1- |C_k(\eta)|^2 \big] \,.\label{unitasupho}\ee Then the factorized form (\ref{ansat}) for superhorizon modes, combined with eqn.
 (\ref{unitasupho}) leads to
 \be F[k;\eta] = \frac{2\pi^2}{(-\eta)^{-2\Delta}}\,\big[ 1- |C_k(\eta)|^2 \big]\,, \label{fketa}\ee and for $-k\eta \gg 1$ and $-p\eta \ll 1$ we find to leading order in $\Delta$
 \be |C_{p}(k;\eta)|^2 = \frac{2\pi^2\,\Delta}{V\,p^3\,(-p\eta)^{-2\Delta}}\,\big[ 1- |C_k(\eta)|^2 \big]\,;\label{Cpetasup}\ee the same result is valid in the region $-k\eta \gg 1$ with $-|\vk-\vp|\eta \ll 1$ by replacing $p \leftrightarrow |\vk-\vp|$.

 The long wavelength limit of eq.(\ref{Cpetasup}) requires a careful treatment. Since $|C_p(\eta)|^2=n_p(\eta)$ is the distribution function of particles, for a fixed volume $V$ there is an infrared divergence in the occupation as $p \rightarrow 0$. However, our goal is to trace over the superhorizon quanta from the decay since the initial conformal time $-\eta_0$ up to conformal time $\eta \rightarrow 0^-$. This entails that the lower momentum cutoff is determined by the mode that just becomes superhorizon at the initial time, namely

 \be p_m = -1/\eta_0 \,. \label{pmin} \ee

 Now the calculation of the entanglement entropy is straightforward: let us consider
 \be I= \sum_{(-1/\eta_0)\leq p \leq (-1/\eta)} |C_{p}(k;\eta)|^2 \ln\Big[ |C_{p}(k;\eta)|^2\Big] \equiv I_1 + I_2 \label{Idef}\ee with
 \bea I_1 & = &  \Big[1-|C_k(\eta)|^2\Big]\,\ln\Big[\frac{ 2\pi^2\Delta (-\eta_0)^3}{V}\,\Big[1-|C_k(\eta)|^2\Big]  \Big]\,\Delta \,\int^{(-1/\eta)}_{(-1/\eta_0)} (-p\eta)^{2\Delta} \frac{dp}{p} \nonumber \\ & = &  \frac{1}{2}\, \Big[1-|C_k(\eta)|^2\Big]\,\ln\Big[\frac{2\pi^2\Delta(-\eta_0)^3}{V}\,\Big[1-|C_k(\eta)|^2\Big] \Big]\, \,\Big[1-x_m^{2\Delta}\Big]   \label{I1}\eea where we have introduced
 \be   x_m = \frac{\eta}{\eta_0} \label{xm}\ee and changing integration variable to $x = -p\eta$
 \bea I_2 & = & - \Big[1-|C_k(\eta)|^2\Big]\,\Delta  \,\int^{1}_{x_m} x^{2\Delta-1} \ln \Big[\frac{x^{3-2\Delta}}{x^3_m} \Big] \, dx \nonumber \\ & = & \frac{1}{2}\,\Big[1-|C_k(\eta)|^2\Big] \Bigg\{3 \ln[x_m]   + \frac{3-2\Delta}{2\Delta}\,\Big[1-(x_m)^{2\Delta}\Big] -2\Delta \,(x_m)^{2\Delta}\,\ln[x_m] \Bigg\}\,. \label{I2} \eea It is now clear that we can set $x_m \rightarrow 0$ safely in $I_1$ and in the terms that \emph{do not feature poles in $\Delta$} in  $I_2$. The terms in $I_2$ that feature the $\ln[x_m]$ and the (single) pole in $\Delta$, namely $(3/2\Delta) \times [1-(x_m)^{2\Delta}]$   yield the leading contribution for $\Delta, x_m  \ll 1$.

 Therefore for $\Delta \ll 1$ and $x_m \ll 1 $ we find for the entanglement entropy to leading order
 \bea S(\eta)  & \simeq &  \frac{\alpha}{(k\eta)^2}\,e^{-\frac{\alpha}{(k\eta)^2}}  - \Big[1-e^{-\frac{\alpha}{(k\eta)^2}}\Big]\,\ln\Big[1-e^{-\frac{\alpha}{(k\eta)^2}}\Big] \nonumber \\ & + \frac{1}{2} &
  \Big[1-e^{-\frac{\alpha}{(k\eta)^2}}\Big]~ \Bigg\{3 \ln\Big[\frac{a_i H_i}{a_0 H_0} \Big]+\ln\Big[ \frac{1}{2\pi^2\,\Delta}\Big]+\frac{3}{2\Delta}\,\Big[Z[\eta]-1+e^{-Z[\eta]} \Big]\Bigg\}\label{entrofinDS}\eea where
  \be Z[\eta] =  {2\Delta} ~ \ln\Big[ \frac{\eta_0}{\eta} \Big] \,,\label{weta}\ee  $\alpha$ is given in eqn. (\ref{sigfina}) and we have set $-\eta_0 = 1/(a_i H_i)$   and $V= 1/(a_0 H_0)$ with $a_i,a_0$ the scale factor, and $H_i,H_0$ the values of the Hubble parameter at the beginning of inflation (i) and today ($0$) respectively, taking the physical volume today to be the Hubble volume, therefore $a_i H_i/a_0 H_0 \simeq 1$.  The function $Z-1+e^{-Z}$ is   manifestly (semi) positive and monotonically increasing,  behaving as $\simeq Z^2/2$ for $Z \ll 1 $ and as $\simeq Z$ for $Z\gg 1$. As $\eta \rightarrow 0$ the entanglement entropy grows monotonically during the time evolution.

 We can $Z[\eta]$ in terms of the number of e-folds since the beginning of inflation as
    \be Z[\eta] \simeq 40\,\frac{M^2}{H^2}\,\big[1 + (N_e(\eta)-N_T)/N_T\big] \,,\label{Wapp} \ee where $N_e(\eta)$ is the number of e-folds during inflation at (conformal) time $\eta$ and $N_T \simeq 60$ is the total number of e-folds of the inflationary stage.

  \section{Wave packets:}
  The discussion above treated the initial and product particles in terms of plane waves. However, given the existence of a horizon and the intricacies that can give rise to for non-localized states, we now generalize the treatment to the case of wave packets. Quantization in a finite volume $V$ is used throughout.   Fock states describing single particle plane wave states of momentum $\vk$,   $|1_{\vk}\rangle$, are normalized such that
\be \langle 1_{\vk}|1_{\vk'}\rangle = \delta_{\vk,\vk'} \label{normalization}\,.\ee
 Localized single particle states are constructed as linear superpositions
\be \overline{| \vk_0,\vx_0\rangle} = \sum_{\vk}C (\vk;\vk_0;\vx_0)\,|1_{\vk}\rangle \label{localsp}\ee where $C_(\vk;\vk_0;\vx_0)$ is the amplitude, normalized so that
 \be \overline{\langle \vk_0,\vx_0} | \overline{\vk_0,\vx_0\rangle} = \sum_{\vk}|C (\vk;\vk_0;\vx_0)|^2=1 \,.\label{normawp}\ee
 For a monochromatic plane wave $C (\vk;\vk_0;\vx_0)= \delta_{\vk,\vk_0}$. The spatial wave function corresponding to the wave packet is given by
 \be \Upsilon(\vec{x}) = \frac{1}{\sqrt{V}} \sum_{\vk}C (\vk;\vk_0;\vx_0)\,e^{-i\vk\cdot\vx} \,. \label{wafu}\ee The normalization (\ref{normawp}) implies
 \be \int d^3 x |\Upsilon(\vx)|^2 =1 \,.\label{normawafu}\ee For a monochromatic plane wave it follows that $\Upsilon(\vx)$ is a volume normalized plane wave.

 The total number of particles  and average momentum of the wave packet are given by
  \be N(\vk_0, \vx_0)= \overline{\langle  \vk_0,\vx_0} |\sum_{\vk}   \,a^{\dagger}_{ \vk} a_{ \vk} |\overline{ \vk_0,\vx_0\rangle} = \sum_{\vk}  |C (\vk;\vk_0;\vx_0)|^2 = 1 \label{numberwp} \ee and
 \be \overline{\langle  \vk_0,\vx_0} |\sum_{\vk} \vk \,a^{\dagger}_{ \vk} a_{ \vk} |\overline{ \vk_0,\vx_0\rangle} = \sum_{\vk}\vk |C (\vk;\vk_0;\vx_0)|^2 \label{kave} \ee respectively, where $a^{\dagger}_{ \vk}; a_{ \vk}$ are the creation and annihilation operators. If $\vk_0$ is identified with the average momentum of the wave packet we assume that
 \be  C(\vk;\vk_0;\vx_0) = C (\vk-\vk_0;\vx_0)\,, \label{propk}\ee and the isotropy of $|C(\vk;\vec{0},\vx_0)|^2$.

As a specific example we consider Gaussian wave packets,
\be   C (\vk-\vk_0;\vx_0) = \Bigg[\frac{8\,\pi^\frac{3}{2}}{\sigma^3\,V} \Bigg]^\frac{1}{2}~e^{-\frac{(\vk-\vk_0)^2}{2\sigma^2}}~e^{i(\vk-\vk_0)\cdot\vx_0} \label{gaussianwf}\,,\ee where $\sigma$ is the localization in momentum space. The spatial wave function is
\be \Upsilon(\vx) = \Bigg[\frac{\sigma}{\sqrt{\pi}}\Bigg]^{3/2}\, e^{-i\vec{k}_0\cdot \vec{x}}\,e^{-\frac{\sigma^2}{2}(\vec{x}-\vec{x}_0)^2} \,.\label{psigau}\ee The spatial wave function is localized at $\vec{x}_0$ with  localization length   $1/\sigma$ and the momentum wave function is localized at $\vec{k}_0$ which is the average momentum in the wave packet  and the momentum localization scale is $\sigma$. The plane wave limit is obtained by formally identifying $\sigma/\sqrt{\pi} \rightarrow 1/V^{1/3}~~;~~V \rightarrow \infty$.

 In terms of these wave functions   the overlap of two wave packets with different momenta localized at different spatial points is
\be \overline{\langle \vq_0;\vx_0}|\overline{ \vk_0;\vx_0\rangle }=  e^{-\frac{(\vk_0-\vq_0)^2}{4\sigma^2}}  \label{overlap}\,. \ee In the limit $\sigma \rightarrow 0$ the overlap becomes a Kronecker delta, and in particular for $k_0,q_0 \gg \sigma$ it follows that the wavepackets are nearly orthogonal since the overlap is non-vanishing for $\Delta k = k_0-q_0 \sim \sigma$ so that $\Delta k/k_0 \ll 1$.

From the identity (\ref{propk}) we can infer the following important property of these wave packets which will be useful below:
\be \sum_{\vk}C (\vk-\vk_0;\vx_0)\,|1_{\vk-\vq}\rangle = |\overline{ \vk_0-\vq;\vx_0\rangle} \,.\label{prop}\ee Although this result is evident with the Gaussian wave packets (\ref{gaussianwf}) it is quite general for localized functions of $\vk-\vk_0$.

The wave packet description is easily incorporated into the Wigner-Weisskopf approach to the description of the full time evolution of the quantum state of the decaying parent particle. The interaction picture quantum state (\ref{expastate}) is generally written as
\be |\Psi(\eta)\rangle = \sum_{\vk} C (\vk,\vk_0;\vec{x}_0;\eta)|1 _{\vk}\rangle  + \sum_{\kappa} \mathcal{C}_{\kappa}(\eta) |\kappa\rangle \label{qstate2}\ee where the states $|\kappa\rangle$ are multiparticle states, with the initial conditions
\be C (\vk;\vk_0;\vec{x}_0;\eta_0) = C (\vk-\vk_0;\vec{x}_0 )~~;~~\mathcal{C}_{\kappa}(t=0) =0 \,, \label{iniwp}\ee where $C (\vk-\vk_0;\vec{x}_0 )$ describe the localized wave packet of the single  particle state at the initial time, for example (\ref{gaussianwf}).

Generalizing the state (\ref{psi}) describing the time evolved state to lowest order in $\lambda$, to a wave packet localized at the origin in space with the gaussian profile (\ref{gaussianwf}),  we can write
\be |\Psi(\eta)\rangle_I = \sum_{\vk} C_1 (\vk-\vk_0;\vec{0};\eta)|1_{\vk}\rangle +\sum_{\vp,\vk}C_2(\vk,\vp,\vk_0;\eta)|1_{\vk-\vp};1_{\vp}\rangle\,, \label{psiwp}\ee with the initial condition
\be C_1 (\vk-\vk_0;\vec{0};\eta_0) = C (\vk-\vk_0;\vec{0})~~;~~ C_2(\vk,\vp,\vk_0;\eta_0) =0 \label{iniconwp}\ee with $C (\vk-\vk_0;\vec{0})$ given by (\ref{gaussianwf}).

Recall that our goal in this article was to obtain the entanglement entropy associated with the decay of single particle states with sub-Hubble physical momenta all throughout the inflationary stage, assuming that near de Sitter inflation lasts a finite time. Namely the physical wavelength of the single particle state is always deep within the Hubble radius during the evolution. A wave packet description of single particle states, therefore must be in terms of wave packets whose physical spatial localization scale  is always much smaller than the Hubble radius. Hence, we will consider wavepackets that are i) sharply localized in comoving momentum with an average momentum $\vk_0$ with $ k_0 \gg H; k_0 \gg \sigma$, the latter condition ensuring a sharp localization around $k_0$,  and ii) with comoving spatial localization scale $1/\sigma  \lesssim  1/H$ so that the wavepacket is localized well within the Hubble radius. Namely the condition for the wavepacket to describe single particle states with a sharp localization in momentum and with spatial localization length scale  smaller than or of the order of the Hubble radius implies the following constraint:
\be k_0 \gg \sigma \gtrsim  H\,.  \label{wpcondi}\ee  Furthermore, consistency in tracing over degrees of freedom with super-Hubble \emph{physical} wavelengths requires that the wavepacket is mainly composed of components with comoving momenta corresponding to physical wavelengths that are always inside the Hubble radius throughout the near de Sitter stage. This condition requires  $-k_0\eta \gg -\sigma\eta \gg 1$ so that components of the wavepacket with super Hubble physical wavelengths are exponentially suppressed.

The Wigner Weisskopf method follows the steps described in detail above. The interaction Hamiltonian connects   the \emph{single particle plane wave states} $|1 _{\vk} \rangle$ with the two-particle plane wave states $|1_{ \vk-\vp};1_{\vp} \rangle$ with matrix elements given by (\ref{mtxele1})
 leading to the set of equations
 \be  \frac{d}{d\eta}C_1 (\vk-\vk_0;\vec{0};\eta) =       -\int^{\eta}_{\eta_0} \Sigma(k,\eta,\eta')\,C_1 (\vk-\vk_0;\vec{0};\eta')\,d\eta' \,, \label{Ckwwwp}\ee
 \be C_2(\vk,\vp,\vk_0;\eta) = -i \int^{\eta}_{\eta_0} \mathcal{M}(p;k;\eta')\, C_1 (\vk-\vk_0;\vec{0};\eta')\, d\eta' \,.\label{coefpkwp}\ee Implementing the Markovian approximation as in the plane wave case with the initial conditions (\ref{iniconwp}) we find
\be  C_1 (\vk-\vk_0;\vec{0};\eta) =  C_1 (\vk-\vk_0;\vec{0};\eta_0)\,C_{k}(\eta) ~~;~~  C_2(\vk,\vp,\vk_0;\eta) =  C_1 (\vk-\vk_0;\vec{0};\eta_0)\, C_p(k;\eta)\,, \label{wwsolswp}\ee where $C_{k}(\eta); C_p(k;\eta)$ are the solutions of the Wigner-Weisskopf equations \emph{for plane waves}, given by (\ref{Ckfinal},\ref{coefpk}).

To obtain the reduced density matrix we would need to carry out the integration over the wavepacket variable $\vk$. The wave packet profile (as function of comoving wavevectors)  is chosen to be sharply peaked at $\vk_0$ with a width $\sigma \ll k_0$. Therefore upon integration we can Taylor expand the integrand around $\vk = \vk_0$ and integrate term by term in the Taylor expansion in $\vk-\vk_0$, because the wavepacket profile is a function of $|\vk-\vk_0|$ it follows that the corrections are a series in $\sigma^2/k^2_0 \ll 1$. An example of a quantity that must be integrated in $\vk$ are the matrix elements (\ref{mtxele1}), which upon being integrated with the wavepacket profile can be simply written as $\mathcal{M}(p;k_0;\eta)+\mathcal{O}(\sigma^2/k^2_0)+\cdots$. The same argument applies to the coefficients \bea C_1 (\vk-\vk_0;\vec{0};\eta) =  C (\vk-\vk_0;\vec{0})\,C_{k_0}(\eta)+\mathcal{O}(\sigma^2/k^2_0)+\cdots \nonumber \\ C_2(\vk,\vp,\vk_0;\eta) =  C  (\vk-\vk_0;\vec{0})\, C_p(k_0;\eta)+\mathcal{O}(\sigma^2/k^2_0)+\cdots \label{expawp}\eea Therefore, to leading order in $\sigma^2/k^2_0$ the reduced density matrix becomes
\bea && \rho_r(\eta)       =  |C_{k_0}(\eta)|^2
\sum_{\vk} \Big( C  (\vk-\vk_0;\vec{0})|1_{\vk} \rangle  \Big)\, \sum_{\vk'}\Big( C^*  (\vk'-\vk_0;\vec{0})  \langle 1_{\vk'}|\Big) + \nonumber \\&&  2   \sum_{(-1/\eta_0) < p <(-1/\eta)}|C_p(k_0;\eta)|^2     \,\sum_{\vk,\vk'}\Big( C  (\vk-\vk_0;\vec{0})|1_{\vk-\vp}\rangle \Big) \,\Big( C^*  (\vk'-\vk_0;\vec{0})
  \langle 1_{\vk'-\vp}|\Big)   \,. \label{rhowp}    \eea

  We emphasize that the trace over the superhorizon modes leading to the reduced density matrix (\ref{rhowp}) has been carried out in the   orthonormal   \emph{plane wave basis}.

  Using the definition of the wavepacket single particle states (\ref{localsp}) and the property (\ref{prop}) we finally find to leading order in $\sigma^2/k_0^2\ll 1 $
  \be \rho_r(\eta) = |C_{k_0}(\eta)|^2 ~ \overline{| \vk_0,\vec{0}\rangle}~\overline{\langle \vk_0,\vec{0}|}+ 2
  \sum_{(-1/\eta_0) < p <(-1/\eta)}  |C_p(k_0;\eta)|^2 ~ \overline{| \vk_0-\vp,\vec{0}\rangle}~\overline{\langle \vk_0-\vp,\vec{0}|}. \label{rhoredwp}\ee

  For $k_0 \gg p,\sigma$ the wave-packet states $\overline{| \vk_0-\vp,\vec{0}\rangle}$ contain plane wave components with subhorizon momenta $\simeq \vec{k}_0 -\vec{p}$ since components with wavevectors that are very different from this value are exponentially suppressed. Therefore these wavepacket states are very nearly plane wave states with subhorizon momenta $k_0  \gg -1/\eta$.

  Therefore, to leading order in $\sigma^2/k^2_0$, the reduced density matrix in terms of the wave packet single particle states features the same form as for the plane wave case with the only modification being the replacement of the single particle Fock states by the localized wavepacket states of single particles.  As a corollary,  to leading order in $\sigma^2/k^2_0$ the entanglement entropy is the \emph{same} either for localized wavepackets or plane waves.

 The logarithmic dependence of the entanglement entropy (\ref{entrofinDS}) on the volume factor has a clear statistical interpretation independent of whether the description is in terms of localized wavepackets or plane wave states. Consider a dilute gas of particles whose statistical distribution  or phase space density is $f_p$. The total \emph{density} of particles is
\be \frac{N}{V} = \int \frac{d^3p}{(2\pi)^3} \, f_p \label{density} \ee and the Von-Neumann entropy of this (dilute) gas is
\be S_{VN} = - \sum_p f_p \ln[f_p]  = - V \int \frac{d^3p}{(2\pi)^3} \, f_p \, \ln[f_p]\,. \label{SVN}\ee  If the number of particles remains finite in the large volume limit, namely if the particle density scales $\propto 1/V$ in this limit, then it follows that $f_p \propto 1/V$. On the contrary,  if $f_p$ is independent of the volume   as in the cases of the Maxwell-Boltzmann, Bose-Einstein or Fermi-Dirac distributions,   the total density is \emph{finite} in the infinite volume limit and the entropy is extensive. For a finite number of particles (vanishing particle density in the infinite volume limit) $f_p \propto 1/V$ and the Von-Neumann entropy is \emph{not} extensive,
\be S_{VN} \propto N \ln[V]\,. \label{nonexSVN}\ee This is \emph{precisely} the origin of the logarithmic dependence on the volume of the entanglement entropy: the initial state has one particle within a Hubble volume and the final state has one (of each) daughter particle, the distribution function of the daughter particles at asymptotically long times after the decay of the parent particle is $|\mathcal{C}_{\chi \psi}(p,\infty)|^2 \propto 1/V$ the inverse volume dependence is the statement that there is a finite number of particles distributed in phase space\footnote{In the first reference in \cite{bala}, only the coupling was kept in the $\ln|C(k)|$ and terms that feature a volume dependence in $|C(k)|$ were discarded as subleading. This explains a discrepancy in the logarithmic volume dependence between this ref. and our results.}. Obviously this volume dependence is independent of whether the states are described by  plane waves or wave packets, but is a statement of the simple fact that the number of particles in the volume $V=(-1/\eta_0)^3$ is finite. The dependence on the scale factor reflects the fact that more
modes are crossing the Hubble radius, but the total number of particles described by these modes is still finite.

  \section{Cascade processes: the way forward}\label{sec:cascade}

  In the previous section we implemented the Wigner Weisskopf method to lowest order in $\lambda^2$, but the method itself is much more general. It relies on a perturbative expansion, a truncation of the hierarchy at a given order in this expansion, and a resummation of the resulting self-energy terms that yield the long time asymptotics. For example, in quantum optics it has been implemented to study the cascade decay of many level atoms\cite{books,book2}.  As shown in\cite{desiter} this resummation is a real time version of the Dyson resummation of self-energies and is equivalent to a dynamical renormalization group resummation of secular terms.

 In this section we set up a roadmap to  study higher order processes and along the way we exhibit the relation between the Wigner-Weisskopf method and the resummation of self-energy diagrams and a diagrammatic expansion. Given the discussion on wavepackets in the previous section, we will restrict ourselves to treating the plane wave case.

\begin{figure}[h!]
\includegraphics[keepaspectratio=true,width=3in,height=3in]{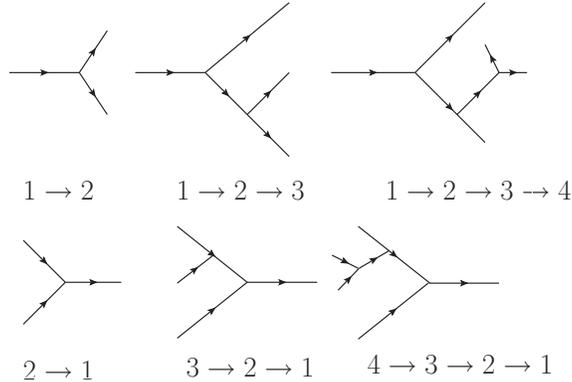}
\caption{Upper diagrams: cascade decay $1\rightarrow 2\rightarrow 3 \rightarrow 4$, each vertex corresponds to a matrix element $M_{ij}\propto \lambda$. Lower diagrams: inverse processes, each vertex corresponds to the matrix element $M_{ji}=M^*_{ij} \propto \lambda$.  Vacuum disconnected diagrams are neglected. }
\label{fig:cascade}
\end{figure}

 The lack of kinematic thresholds in inflationary cosmology implies that the decay of quanta occur in a \emph{cascade} process. For example with a cubic interaction as studied above, a state with a single quanta can decay into a state with two other quanta, in turn each one of the quanta in this state can decay into two other quanta, therefore a single particle state will decay via a ``cascade'': $1\rightarrow 2 \rightarrow 3 \rightarrow 4 \cdots$ depicted in fig.(\ref{fig:cascade}).

Each branch of the cascade corresponds to an interaction vertex and another power of the coupling, showing that the branches of the cascade are suppressed in perturbation theory. For example, the amplitude for $3$ particles is down by a factor of $\lambda$ (trilinear coupling) with respect to the two particle one, the four particle state is suppressed by another power of $\lambda$, etc.

   To simplify notation, let us define matrix elements that connect a state of $i$ quanta with a state of $j$ quanta via the interaction Hamiltonian $H_I$,
   \be M_{ij}(\eta) = \langle[i]|H_I(\eta)|[j]\rangle \propto \lambda \,.\label{matele}\ee Here $[i],[j]$ describes the set of $i,j$ quanta with different values of momenta. As studied above, we see that a state with a single quanta of (comoving) momentum $\vec{k}$ is connected via $H_I$ to a state with two quanta, with momenta $\vec{q},\vec{k}-\vec{q}$ respectively. The matrix element for this process is $\langle[1]|H_I(\eta)|[2]\rangle =  \langle 1_{\vk}|H_I(\eta)|1_{\vec{q}};1_{\vec{k}-\vec{q}}\rangle $ where the set of values $\vec{q}$ defines the two-quanta states $[2]$. Thus the generic matrix elements between single quanta states and two quanta states in this set are $ \langle[1]|H_I(\eta)|[2]\rangle   \equiv M_{12}(\eta) \propto \lambda $. The \emph{inverse} process $[j]\rightarrow [i]$ is described by the matrix element $\langle [j] | H_I(\eta) |[i]\rangle = M_{ji}(\eta) = M^*_{ij}(\eta)$ because the Hamiltonian is hermitian.

 In what follows we will only consider \emph{connected} diagrams or processes, neglecting disconnected diagrams which do not describe transitions but rather a renormalization of the vacuum state (for discussions see\cite{desiter}). Consider the cascade decay of a single particle state $|1_{\vk}\rangle$ into three particles, along with their inverse processes, neglecting the disconnected (vacuum) diagrams  the typical sequence is shown in fig. (\ref{fig:cascade}) and the quantum state is given by
   \be |\Psi(\vk,\eta)\rangle = C_1(k,\eta) |1_{\vk}\rangle + \sum_{\vp} C_2(\vk,\vp;\eta) |1_{\vk-\vp};1_{\vp}\rangle + \sum_{\vp,\vq} C_3(\vk,\vp,\vq;\eta) |1_{\vk-\vp};1_{\vq};1_{\vp-\vq}\rangle + \cdots \label{state3pars}\ee

   The set of Wigner-Weisskopf equations are obtained straightforwardly as in the previous section. An important aspect in obtaining these equations is that a particular state with $n$ particles with a fixed set of momenta has branched out from one ``ancestor state'', whereas it branches forward into an $n+1$ particle state where the new particle has an arbitrary momentum that is summed over. As an example of this pattern consider the $3$ particle state $|1_{\vk-\vp};1_{\vq};1_{\vp-\vq}\rangle$ for a fixed value of $\vp$ and $\vq$ (the value of $\vk$ is fixed by the initial state). This state branched out from the two particle state $|1_{\vk-\vp};1_{\vp}\rangle$ (up to relabelling the momenta and indistinguishability of the particle states), therefore \emph{it only has one ``ancestor''} as a consequence of momentum conservation. However, it branches out to $4$ particle states of the form $|1_{\vk-\vp};1_{\vq};1_{\vec{l}};1_{\vp-\vq-\vec{l}}\rangle$ where the wavector $\vec{l}$ must be summed over.

The hierarchy of Wigner-Weisskopf equations reads in shortened  notation
\bea \dot{C}_1(\eta) &  =  & -i \sum_{[2]} M_{12}(\eta) C_{[2]}(\eta) \label{c1ww}\\
\dot{C}_2(\eta) &  =  & -i M_{21}(\eta) C_1(\eta) -i \sum_{[3]} M_{23}(\eta) C_{[3]}(\eta) \label{c2ww} \\
\dot{C}_3(\eta) &  =  & -i   M_{32}(\eta) C_{2}(\eta) -i \sum_{[4]} M_{34}(\eta) C_{[4]} (\eta) \label{c3ww} \\
\vdots & = & \vdots \nonumber \eea The labels  without brackets in the coefficients $C_n$ correspond to a particular state of $n-particles$ with a fixed set of momenta compatible with total momentum conservation, whereas the sums over $[n]$ are over the n-particle states compatible with the set of wavenumbers determined by momentum conservation. The terms shown in the hierarchy (\ref{c1ww},\ref{c2ww},\ref{c3ww}) are the ones depicted in fig. (\ref{fig:cascade}) and their inverse processes: if the Hamiltonian connect the states $[i]$ with the states $[j]$
it also connects $[j]$ back with $[i]$, these are the inverse processes depicted in fig. (\ref{fig:cascade}).

The two  terms in eqns. (\ref{c2ww},\ref{c3ww}) have an illuminating interpretation. The first terms correspond to the ``population gain'' of the states with two and three particles from the decay of their ancestors states with one and two particles  respectively, while the second terms represent the ``loss'' or decay of the amplitudes into states with one more particle. Because of the initial conditions $C_1(\eta_0) =1;C_{n\neq 1}(\eta_0) =0$, it follows that $d{|C_2|^2}/d\eta \propto \lambda^2 + \lambda^3+\cdots ~;~d{|C_3|^2}/d\eta \propto \lambda^4 + \lambda^5 +\cdots$ so that the (conformal) time dependence of the coefficients also follows a hierarchy: the three particle state ``fills up'' on time scales  $\propto 1/\lambda^2$ larger than the two particle state, the four particle state on time scales $\propto 1/\lambda^2$ larger than the three particle state, etc.

Let us consider truncating the hierarchy beyond the three particles intermediate state, namely set $C_{[4]} = C_{[5]} = \cdots =0$ along with all the other higher terms in the hierarchy. We then proceed to solve the equations from the bottom up with the initial conditions $C_1(\eta_0) =1;C_{[2]}(\eta_0)=C_{[3]}(\eta_0) = \cdots =0$. We obtain
\bea C_3(\eta) &  = &  -i \int^{\eta}_{\eta_0}   M_{32}(\eta') C_{2}(\eta')\,d\eta'  \label{c3sol} \\ \dot{C}_2(\eta) &  = &  -i     M_{21}(\eta) C_1(\eta)     -    \int^{\eta }_{\eta_0} d\eta_1 \sum_{ [3]} M_{23}(\eta ) M_{32}(\eta_1 )\,C_{2}(\eta_1)\,.   \label{c2sol} \eea

The first term in (\ref{c2sol}) describes the build-up of the two-particle amplitude from the decay of the initial single particle state, whereas the second term describes the decay of the two-particle state into three particles via the cascade decay. Since the matrix elements are $\propto \lambda$ we can solve eqn. (\ref{c2sol}) iteratively in perturbation theory up to the order considered in the hierarchy, in order to understand the time scales,
\bea  C_2(\eta) &  = &  -i \int^{\eta}_{\eta_0}   M_{21}(\eta_1) C_1(\eta_1)\,d\eta_1 \nonumber \\ & + i &  \int^{\eta}_{\eta_0} d\eta_1 \int^{\eta_1}_{\eta_0} d\eta_2 \int^{\eta_2}_{\eta_0} d\eta_3 \sum_{ [3]} M_{23}(\eta_1) M_{32}(\eta_2)M_{21}(\eta_3)\,C_{1}(\eta_3)+ \cdots \,.   \label{c2solit} \eea

To make the arguments clear, let us consider Minkowski space time and early time scales so that $C_1 \simeq 1$. Then the two particle amplitude builds up $\propto \lambda t$ (with rate $\propto \lambda$),  and from eqn. (\ref{c3sol}) we see that the three particle state builds up $\propto \lambda^2 t^2 \ll \lambda t$, clearly reflecting that the population of the three particle state builds up much slower than that of the two particle state etc.

The build-up and decay integrals feature secular growth as $\eta \rightarrow 0$ (long cosmic time), and the second step in the Wigner-Weisskopf method provides a non-perturbative resummation of these processes: writing (\ref{c2sol}) as an integro-differential equation
\be \dot{C}_2(\eta) + \int^{\eta}_{\eta_0} \Sigma_{(2)}(\eta,\eta_1) C_2(\eta_1) d\eta_1  = -i     M_{21}(\eta) C_1(\eta) ~~;~~\Sigma_{(2)}(\eta,\eta_1) = \sum_{ [3]} M_{23}(\eta ) M_{32}(\eta_1 )\,, \label{difeqC2}\ee and introducing the Markovian approximation as in eqn. (\ref{Wo}-\ref{marko1})  (the second approximation in the Wigner-Weisskopf method) we find
\be C_2(\eta) = -i  e^{-\gamma_2(\eta)}\,\int^{\eta}_{\eta_0}  M_{21}(\eta_1) C_1(\eta_1) e^{\gamma_2(\eta_1)} \,d\eta_1 ~~;~~ \gamma_2(\eta) = \int^{\eta}_{\eta_0} \Sigma_{(2)}(\eta,\eta')d\eta' \,. \label{C2solww}\ee  This compact expression reveals at once the build-up of the amplitude from $C_1$ and the eventual decay of the two-particle state encoded in $\gamma_2(\eta)$.

A simple perturbative expansion of this expression up to $\mathcal{O}(\lambda^4)$ reproduces (\ref{c2solit}) consistently with the Markovian approximation.

\begin{figure}[h!]
\includegraphics[keepaspectratio=true,width=4in,height=4in]{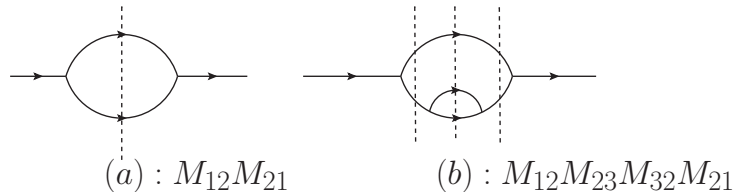}
\caption{The contributions to $C_1$ showing the one and two loop contributions to the self-energy. The dashed lines represent intermediate states with two or three particles, corresponding to the matrix elements $M_{12};M_{21}$ in (a), and similarly for (b). There are other two loop diagrams not shown.  }
\label{fig:twoloops}
\end{figure}

 The last step is to insert this solution into (\ref{c1ww}), solve the integro-differential equation for $C_1$ and insert this  solution into (\ref{c2solit}) and (\ref{c3sol}) respectively. Obviously this procedure leads to a very complicated expression that is not very illuminating. However progress can be made by introducing the perturbative solution (\ref{c2solit}), leading to the following integro-differential equation for $C_1$:
\bea  \dot{C}_1(\eta) &  =  &   -\int^{\eta}_{\eta_0} \sum_{[2]} M_{12}(\eta)M_{21}(\eta_1) C_{1}(\eta_1) \nonumber \\ & + & \int^{\eta}_{\eta_0} d\eta_1 \int^{\eta_1}_{\eta_0} d\eta_2 \int^{\eta_2}_{\eta_0} d\eta_3 \sum_{[2],[3]}M_{12}(\eta) M_{23}(\eta_1) M_{32}(\eta_2)M_{21}(\eta_3)\,C_{1}(\eta_3)\  \label{finic1}\eea The first and second terms have a simple interpretation in terms of one and two loop self-energies as depicted in fig. (\ref{fig:twoloops}) (only one two loop contribution is shown).

The dashed lines cut through multiparticle states and indicate similar rules to the Cutkosky rules of quantum field theory that relate the absorptive parts of self-energy diagrams to intermediate multiparticle states.

In order to make progress in the solution of (\ref{finic1}) the second part of the Wigner-Weisskopf method invokes a Markovian approximation, just as that described in section (\ref{subsec:ww}) implemented to lowest order. This approximation is again justified in a weak coupling expansion and is the statement that $\eta$ derivatives of the coefficients are ``slow'' and can be systematically expanded perturbatively. The procedure follows the steps  described by  eqns. (\ref{Wo}-\ref{markoeA}), integrating by parts the kernels of the integrals and keeping consistently up to $\mathcal{O}(\lambda^4)$ we find
\be \dot{C}_1(\eta) + W(\eta) C_1(\eta) = 0 ~~;~~ C_1(\eta_0) = 1 \,, \label{fineqc1} \ee where
\bea W(\eta) & = &  \int^{\eta}_{\eta_0} \sum_{[2]} M_{12}(\eta)M_{21}(\eta_1) d\eta_1 \Big[1 +
 \int^{\eta}_{\eta_0}\int^{\eta_1}_{\eta_0} \sum_{[2]} M_{12}(\eta)M_{21}(\eta_2) d\eta_1 d\eta_2 \Big] \nonumber \\ & + & \int^{\eta}_{\eta_0} d\eta_1 \int^{\eta_1}_{\eta_0} d\eta_2 \int^{\eta_2}_{\eta_0} d\eta_3 \sum_{[2],[3]}M_{12}(\eta) M_{23}(\eta_1) M_{32}(\eta_2)M_{21}(\eta_3)\,. \label{W4}\eea The second term in the bracket in the first line arises from the derivative expansion of the term with the one-loop self-energy (see eqn. (\ref{marko1})); in ref.\cite{desiter} this term is identified as a contribution to wave function renormalization. Therefore
 \be C_1(\eta) = e^{-\int^{\eta}_{\eta_0} W(\eta')d\eta'}\,.
  \label{c1etafi}\ee This expression provides a non-perturbative resummation of self-energies in real time up to two loops and includes the decay of the initial state into intermediate states with two and three particles.

 In Minkowski space time, the initial state decays as $\propto e^{-\Gamma t}$ with $\Gamma = \lambda^2 \gamma_{2}+ \lambda^4 \gamma_{3}+\cdots$ corresponding to the contribution to the self energy from the two particle intermediate states (one loop) three particle intermediate states (two loops) etc, highlighting that the probability of production of the two particle intermediate state occurs on a time scale $\propto 1/\lambda^2$, that of the three particle intermediate state on $\propto 1/\lambda^4$ etc.

  Clearly the decay into two particle states occurs on shorter time scales as this process corresponds   to the one-loop diagram, whereas decay into three particles occurs on much slower scales at this process corresponds to the two-loop contributions.

  It remains to insert this solution into (\ref{c2solit}) and in turn insert the solution for $C_{[2]}$,into (\ref{c3sol}). Because the matrix elements $M_{ij} \propto \lambda$ it follows that if we take $C_{1} \propto \mathcal{O}(\lambda^0)$, then $C_{[2]} \propto \lambda~;~C_{[3]}\propto \lambda^2 \cdots$. The quantum state obtained from the decay of a quanta with momentum $\vk$ is given by
  \be |\Psi(\vk,\eta) \rangle = C_1(\eta) |1_{\vk}\rangle + \sum_{[2]} C_{[2]}(\eta) |[2]\rangle + \sum_{[3]}C_{[3]}(\eta)|[3]\rangle + \cdots \,, \label{statecasca}\ee The states $|[2]\rangle = |1_{\vp};1_{\vk-\vp}\rangle$ and $|[3]\rangle = |1_{\vp_1};1_{\vp_2}:1_{\vk-\vp_1-\vp_2}\rangle$ and the sums over $[2],[3]$ are over $\vp$ and $\vp_1, \vp_2$ respectively.

  Thus the probability of a given two particle state is given by $|C_{2}(\eta)|^2 \propto \lambda^2$, of a given three particle state is $|C_{3}(\eta)|^2 \propto \lambda^4$, etc. This is exactly as in the case of multiphoton processes in quantum electrodynamics or of an atomic cascade of a multilevel atom. Each photon in the final state is associated a probability that is proportional to $\alpha_{em}$, so multiphoton processes are suppressed by powers of the fine structure constant. In this case multiparticle final states are suppressed by powers of $\lambda^2$ for each extra particle in the final state. This is also the case in atmospheric air showers where very energetic particles decay via a cascade process where each branch of the cascade is down by a power of the coupling to the respective channel.

  In the case of cascade decay in Minkowski space time, the probability of finding particles from a particular decay channel is given by the branching ratio of such channel $\Gamma_c/\Gamma_{tot}$, namely ratios of different powers of the couplings. Our result obviously entails the same physics: the probability of a state with three quanta is suppressed by $\lambda^2$ with respect to that with only two quanta, etc.

 Furthermore, the explicit form of $W(\eta)$ (\ref{W4}) clearly shows the separation of time scales: the decay into two particles involves time scales $\propto 1/\lambda^2$ and is determined by the product of matrix elements $M_{21} M_{12}$ whereas the time scales for decay into three particle states is determined by the last term in (\ref{W4}) which implies time scales $\propto 1/\lambda^4$. Therefore there is a hierarchy both in the probability of multiparticle states and the time scales associated with their production from the decay of the parent particle. The cascade decay processes are controlled by the small coupling $\lambda$.

  The entanglement entropy can now be calculated by obtaining the reduced density matrix by tracing over the superhubble degrees of freedom in the pure state density matrix $|\Psi(\vk,\eta) \rangle \langle \Psi(\vk,\eta) |$ and is a straightforward implementation of the steps described in the previous section with the technical complication of the integration over the super Hubble subset of momenta in the multiparticle contributions. This is only a technical difficulty but not a conceptual roadblock, since the contribution to the entanglement entropy from higher multiplicity states will be suppressed by high powers of the coupling $\lambda$. An illustrative example in Minkowski space time is the cascade decay
  $\pi^- \rightarrow \mu^-\,\overline{\nu}_\mu~;~\mu^-\rightarrow e^-\,\overline{\nu}_e\,\nu_\mu$: whereas the pion decays on a time scale $\simeq 2.8 \times 10^{-8}\,\mathrm{secs}$ the muon decays on a time scale $\simeq 2.2 \times 10^{-6}\,\mathrm{secs}$ therefore during a long time interval $10^{-8}\, \mathrm{secs} \leq t \leq 10^{-6}\,\mathrm{secs}$ the two particle state $|\mu^-,\overline{\nu}_\mu\rangle$ yields the largest contribution to the quantum state.

  Furthermore, the unitarity relation (\ref{unione}) entails that
  \be |C_1(\eta)|^2 + \sum_{[2]}|C_{[2]}(\eta)|^2 + \sum_{[3]}|C_{[3]}(\eta)|^2 + \cdots =1 \,, \label{unicasca}\ee which was confirmed in the previous section to leading order in the coupling and $\Delta$.

  In summary: the cascade decay is controlled by the perturbative nature of the interaction, the probability for multiparticle states being suppressed by powers of the coupling constant and the time scales associated with the formation of multiparticle states widely separated by larger powers of $1/\lambda$. Furthermore, in the case under consideration here, the physical momentum of the initial state  is taken to remain deep inside the Hubble radius at all times during inflation. At any large but fixed (conformal time) the initial state maintains a small but non-vanishing population, a two particle state being populated with probability $\lambda^2$, a given three particle state with probability $\propto \lambda^4$ etc. Therefore if the quasi-de Sitter inflationary stage lasts a finite (say $\simeq 60$) number of e-folds, the quantum state will be a linear superposition of many particle states and unitarity implies that each state features a perturbatively small population. An interesting and conceptually puzzling situation arises in the case of eternal de Sitter, since in this case, at asymptotically long times all states would have decayed to vanishing probability in clear contradiction with unitarity, but in this case all physical momenta eventually also become superHubble. Perhaps this puzzling aspect is related to the intriguing results of ref.\cite{polya} and deserves to be studied further.

  While we have established a roadmap and a ``proof of principle'' of the method, undoubtedly there are several aspects that merit a deeper study such as infrared enhancement from superhorizon modes, the issue of unitarity in eternal de Sitter, the detailed aspects of the (conformal) time dependence of the amplitudes of multiparticle states etc. We postpone  the   study of these more technical details of the higher order processes to a future article.

\section{Possible relation to Non-Gaussianity.}

\begin{figure}[h!]
\begin{center}
\includegraphics[height=3in,width=3in,keepaspectratio=true]{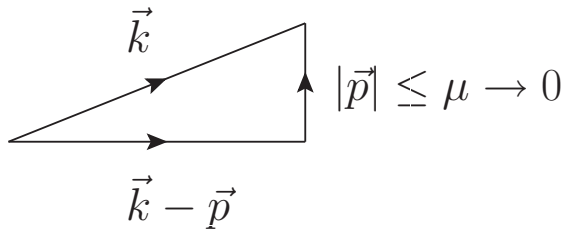}
\caption{Triangle of momenta for the bispectrum (see eqn.(\ref{bispec})) integration over $|\vec{p}|< \mu \rightarrow 0$ corresponds to the highly squeezed limit and yields the pole in $\Delta$.}
\label{triangle}
\end{center}
\end{figure}

   The cubic interaction vertex   suggests a relation between the decay amplitude (see fig.(\ref{fig:decay})) and the non-gaussian bispectrum which is the three point function of the field. The relationship with the  bispectrum becomes more clear by introducing  $G(k,\eta,\eta') = g^*_\nu(k;\eta)g_\nu(k;\eta')$ from which it follows that
\be \int^\eta_{\eta_0} \Sigma(k,\eta,\eta')~d\eta' \propto \int \frac{d^3p}{(2\pi)^3} ~\int \frac{d\eta'}{H\eta'} ~ G(k,\eta,\eta')\,G(p,\eta,\eta')\,G(|\vec{k}-\vec{p}|,\eta,\eta')\,. \label{bispec}\ee The imaginary part of the $\eta'$-integral is proportional to  the \emph{bispectrum} of the scalar field\cite{nongauss,malda}. The main difference is that the self-energy is the integral over one of the momenta. In particular the leading order in $\Delta$, namely the contribution from the infrared enhanced, superhorizon modes, is determined by the highly squeezed limit shown in fig. (\ref{triangle}), which corresponds to the local limit of the non-gaussian correlator.

This connection highlights that this local limit is describing \emph{correlations} between subhorizon and superhorizon modes, these are the correlations that yield the entanglement entropy upon tracing over the superhorizon degrees of freedom.

There are important differences between the scalar field theory with cubic interaction studied here, and the cubic interactions of curvature perturbations in the theory of non-gaussian fluctuations\cite{nongauss,malda}, the main difference being both spatial and (conformal) time derivatives in the interactions. However, the study of ref.\cite{picon} found  that transition probabilities of curvature perturbations (in single field slow roll inflation) are suppressed by slow roll parameters but enhanced by infrared logarithms, which are similar to those emerging in the $\Delta \rightarrow 0$ limit in our study (corresponding to massless fluctuations), thus suggesting that the results obtained above \emph{may} apply to the decay of curvature perturbations and superhorizon entanglement and concomitant entanglement entropy.

\section{Conclusions and further questions}

In inflationary cosmology all particle states \emph{decay} as a consequence of the lack of a global time-like Killing vector which would in turn enforce kinematic thresholds. In this article we have studied the entanglement entropy from the decay of single particle states during de Sitter inflation in a  theory of a light scalar field  with $M^2\ll H^2$ and cubic interactions. The quantum state that describes the single particle decay and the produced particles is a two-particle state entangled by momentum conservation. We have extended and generalized the Wigner-Weisskopf method used in the treatment of spontaneous decay of atomic states to the realm of quantum field theory in an expanding cosmology, and implemented this method to obtain the quantum state that describes the decay of the parent particle and the production of the daughter particles. We showed in detail that this non-perturbative approximation is \emph{manifestly unitary}. The amplitudes for the two-particle entangled state features infrared enhancements that are manifest as poles in $\Delta = M^2/3H^2$  as a consequence of the emission of superhorizon quanta  and we implement a consistent expansion in $\Delta$ to leading order to obtain the (pure state) density matrix that describes the decay of the parent and production of daughter particles. When the parent particle's wavelength is inside the horizon, the density matrix elements for the produced particles are dominated by the contribution of \emph{superhorizon} momenta of one of the daughter particles, describing entanglement, correlation and coherences across the horizon.
Tracing the pure state density matrix over the superhorizon modes we obtain a \emph{mixed state} density matrix from which we calculate the Von Neumann entanglement entropy, which describes the loss of information from the correlations between sub and superhorizon modes due to the non-observation of these latter states.
We find that the entanglement entropy is  enhanced in the infrared by a factor of  $\ln[1/\Delta]$ and grows logarithmically with the physical volume as a consequence of more modes crossing the Hubble radius during the inflationary stage.

The generalization to the description of single particle states in terms of wavepackets spatially localized within the Hubble radius but localized in momentum was provided. Under the conditions that the average wavevector of the wave packet be associated with subHubble wavelengths all throughout the near de Sitter stage, we showed the equivalence between the plane wave and wave packet description and assessed the corrections in terms of the ratio of the width of the wavepacket in momentum space and the average momentum associated with the single particle state.

The lack of kinematic thresholds implies that particle decay occurs in a cascade process, namely $1\rightarrow 2 \rightarrow 3 \cdots$. We have extended the  Wigner-Weisskopf method to establish a framework to study the cascade decay and analyzed in detail the process up to a three particle branching in the cascade, but the results are quite general. We showed that for weak coupling (here we considered a cubic coupling) the probability of multiparticle states is suppressed by powers of the coupling, for example in the case of cubic coupling the three particle state is suppressed by $\mathcal{O}(\lambda^2)$ with respect to the two particle state, the four particle $\mathcal{O}(\lambda^4)$ etc. We have established a relation between the different multiparticle processes and higher order loop contributions in the self-energy, just as in the case of Cutkosky rules in Minkowski space-time. This relation clearly shows that just as the probability of higher multiparticle states is suppressed by high powers of the coupling, the time scales for decay into higher multiplicity states are widely separated by inverse powers of $\lambda^2$. Therefore the cascade decay is controlled by the weak coupling, just as multiphoton processes in QED.

\textbf{Further questions:}

\vspace{1mm}

 \textbf{a):} An important feature of inflationary cosmology is that physical wavelengths that cross the Hubble radius during inflation, re-enter the Hubble radius (now the particle horizon) during radiation or matter domination and these quantum fluctuations are the seeds of temperature anisotropies and inhomogeneities.

The entanglement entropy that we have studied is a measure of the correlations between the entangled subhorizon and superhorizon degrees of freedom as a consequence of interactions, which brings the question of whether upon re-entry the fluctuation modes that were superhorizon during inflation ``bring back'' the quantum correlations and if so how are these manifest in the power spectrum of the CMB? Furthermore, going from quantum fluctuations of the curvature (or gravitational potential) to temperature fluctuations entails replacing quantum averages by statistical averages. Thus it is a relevant question whether this statistical averaging includes the quantum correlations from entanglement. Last but not least, if the quantum states can decay,  it is conceivable that the lack of power in the low multipoles which is present in the cosmological data and has been persistent in the statistical analysis of WMAP7\cite{wmap7},  WMAP9\cite{wmap9} and  Planck\cite{planck} which reports a power deficit at low multipole with $2.5-3\,\sigma$ significance and a recent statistical analysis of the combined dataset\cite{gruppuso}, may be due to the decay of the quantum fluctuations during the inflationary stage. Our study applies to a scalar field in de Sitter space time and in order to answer this question the analysis presented here must be applied to the case of scalar perturbations.

\vspace{1mm}

\textbf{b:)} Furthermore, and in relation with the question above, it is a tantalizing possibility that the superhorizon correlations become manifest as intensity correlations leading to interference phenomena akin to the Hanbury-Brown-Twiss effect discussed in ref.\cite{masimo}.

\vspace{1mm}

 \textbf{c):} how the infrared enhancements modify the higher order multiparticle processes in the cascade decay, and how unitarity is manifest in the (formal) case of eternal de Sitter inflation.

 While at this stage we do not \emph{yet} see a clear observational consequence of the entanglement entropy beyond the theoretical conceptual aspect of information loss from the correlations and superhorizon entanglement, the exploration of potential observational consequences along with the  questions raised above are worthy of further and deeper study, on which we expect to report in the future.

 This study of the \emph{superhorizon entanglement entropy} from particle decay bridges two concepts previously explored in the literature: the entanglement between spatially separated but correlated regions, in our case the correlations between sub and superhorizon quanta of the daughter particles, akin to the superhorizon correlations studied in ref.\cite{maldacena}, and the momentum-space entanglement studied in ref.\cite{bala,minko}. In our study the entanglement entropy is a result of \emph{both} types of concepts, linked together by the interactions but with the distinct aspect of being a non-equilibrium process as a consequence of the cosmological expansion.

\acknowledgments D.B. is deeply indebted to  A. Daley,  and D. Jasnow for  enlightening comments and discussions, L. L. and D.B.  acknowledge partial support from NSF-PHY-1202227. R.~H. was supported in part by the Department of Energy under grant DE-FG03-91-ER40682. He would also like to thank the Cosmology group at UC Davis for hospitality while this work was in progress.

\end{document}